\documentclass{PoS}

\newcommand{\MS}{{\overline{\rm MS}}}
\newcommand{\csw}{c_{sw}}

\title{
\vspace{-26mm}
\begin{flushright}{\normalsize \rm DESY-15-213\\ 
                           Edinburgh 2015/24  \\ 
                           Liverpool LTH 1064 \\ \vspace{-3mm}
                           Adelaide ADP-15-44/T946 }
\end  {flushright}\vspace{4.5mm}
Improving the lattice axial vector current}

\ShortTitle{Improving the lattice axial vector current}

\author{R.~Horsley$^{a}$, Y.~Nakamura$^{b}$, H.~Perlt$^{c}$, P.~E.~L.~Rakow$^{d}$, G.~Schierholz$^{e}$,
\speaker{A.~Schiller}$^{c}$ and
J.~M.~Zanotti$^{f}$
\\
$^{a}$ School of Physics and Astronomy, University of Edinburgh, Edinburgh EH9 3JZ, UK
\\
$^{b}$ RIKEN Advanced Institute for Computational Science, Kobe, Hyogo 650-0047, Japan
\\
$^{c}$ Institut f\"ur Theoretische Physik, Universit\"at Leipzig, PF 100 920, D-04009 Leipzig, Germany
\\
$^{d}$ Theoretical Physics Division, Department of Mathematical Sciences, University of Liverpool,
\\
\hspace{2mm} Liverpool L69 3BX, UK
\\
$^{e}$ Deutsches Elektronen-Synchrotron DESY, 22603 Hamburg, Germany
\\
$^{f}$ CSSM, Department of Physics, University of Adelaide, Adelaide SA 5005, Australia
\\
E-mail: \email{Arwed.Schiller@itp.uni-leipzig.de}
}

\abstract{
For Wilson and clover fermions traditional formulations of the axial vector current
do not respect the continuum Ward identity which relates  the divergence
of that current to the pseudoscalar density.
Here we propose to use a point-split or one-link axial vector current whose divergence exactly 
satisfies a lattice Ward identity, involving the pseudoscalar density
and a number of irrelevant operators.
We check in one-loop lattice perturbation theory with  SLiNC fermion and gauge plaquette action 
that this is indeed the case including order $O(a)$ effects.
Including these operators the axial Ward identity remains renormalisation invariant.
First preliminary  results of a nonperturbative check of the Ward identity are also presented.
}

\FullConference{The 33rd International Symposium on Lattice Field Theory\\
                14 -18 July  2015\\
                Kobe International Conference Center, Kobe, Japan}

\begin{document}

\section{Introduction}

The axial (nonsinglet) vector current defined as
\begin{eqnarray}
  A_{\mu}(x) = \bar \psi(x) \gamma_\mu \gamma_5 \psi(x)
\end{eqnarray}
is an important quantity to study hadronic structures such
as quark masses and meson decay constants.
In the continuum that current satisfies the axial Ward identity (WI)
\begin{eqnarray}
   \partial_\mu A_\mu(x) = 2 m P(x)\,, \quad
   P(x)=\bar  \psi(x) \gamma_5 \psi(x)
\end{eqnarray}
where $P(x)$ denotes the pseudoscalar current and $m$ is the quark mass.

In most lattice studies  with Wilson and clover fermions the  local  lattice
axial vector $A_\mu^{\rm loc}$ plus $O(a)$ improvement is used. 
Knowing the nonperturbative improvement coefficients $b_A$ and $c_A$ and the renormalisation constant $Z_{A^{\rm loc}}$
the renormalised current
\begin{eqnarray}
  A_\mu^{\rm loc,\MS} = Z_{A^{\rm loc}} (1 + b_A a m) \big[A_\mu^{\rm loc} + a c_A \partial _\mu P \big]^{\rm lat}
\end{eqnarray}
has been used to determine e.g. the nucleon axial charge $g_A$ as a benchmark test.
For the considered lattice fermions most of  $g_A$ are 
slightly below the experimental value~\cite{Constantinou:2014tga}. 
A possible reason might be that 
the (continuum) Ward identity is spoiled by lattice artefacts
using such a  local axial vector.

Here we propose to 
use a point-split or one-link
axial vector current  whose divergence  exactly satisfies a lattice WI, involving
the pseudoscalar density and a number of irrelevant operators. 
Operators fulfilling lattice Ward identities and being $O(a)$ improved could remove lattice artefacts 
and simplify renormalisation issues.
Such operators naturally appear in derivation of lattice Ward identities~\cite{Bochicchio:1985xa,Reisz:2000dt}
(for a recent discussion on related issues, see e.g.~\cite{Bhattacharya:2015rsa}).

In~\cite{Reisz:2000dt} a proof of such a WI for Wilson fermions with arbitrary Wilson coefficient $r$
has been performed in 
lattice perturbation theory. Using the plaquette gauge action, the renormalisation factor for 
the point-split axial vector
has been calculated as
\begin{eqnarray}
  Z_{A^{\rm split}}(r)=1 + C_F g^2 \, \xi(r)\,, \quad   16 \pi^2 \,\xi(1)=- 8.663 \,.
\end{eqnarray}

In this contribution we report on a study
of such an axial WI using clover fermions and check it both perturbatively and nonperturbatively.

\section{Lattice axial Ward identity for the point-split axial vector current}

To derive a lattice Ward identity for the point-split axial vector current we start
from an identity on every configuration ($i,j,k,x$ - site positions, no implicit summation)
\begin{eqnarray}
 M^{-1}_{jx} \, \gamma_5 \,\delta_{xi} +\delta_{jx} \, \gamma_5 \, M^{-1}_{xi}
   =
   \sum_k \left(M^{-1}_{jx} \, \gamma_5 \, M_{xk} M_{ki}^{-1}+ M^{-1}_{jk} M_{kx}\, \gamma_5 \, M^{-1}_{xi}\right)\,,
   \label{eq:ident}
\end{eqnarray}
where $M_{ij}$ denotes the fermion matrix. In the case of clover fermions it can be split up into the anticommuting with $\gamma_5$
(the $\gamma_\mu$ terms) part $D$, the Wilson term $W$, the clover term $C$ and a bare mass $m_B$ vanishing for $\kappa=1/8$
\begin{eqnarray}
  M_{ij}= D_{ij}+ W_{ij} +C_{ij} + m_B \delta_{ij} I \,, \quad a m_B= \frac{1}{2 \kappa}-4  \,.
  \label{eq:Msplit}
\end{eqnarray}
Averaging (\ref{eq:ident}) over gauge configurations, using the splitting (\ref{eq:Msplit})
and rearranging the terms, 
we get
\begin{eqnarray}
  {G_{jxi}[\partial \cdot A]= 2 m_B G_{jxi}[P]} 
  {- S_{jx} \gamma_5 \delta_{xi} -\delta_{jx} \gamma_5 S_{xi}}
  + 2 G_{jxi}[O_C]  {+G_{jxi}[O_W]} \,.
  \label{eq:latWI}
\end{eqnarray}
Here
$S_{jx}$ is the fermion propagator (two-point function)
and
$G_{jxi}[O]$ denotes the fermion-line connected three-point function of operator $O$.
The different operators are of the form
\begin{eqnarray}
  a \, \partial \cdot A&=&\sum_\mu[A_\mu^{\rm split}(x) -A_\mu^{\rm split}(x-a \hat \mu) ]
  \,, \nonumber \\
A_\mu^{\rm split}(x)&=& \frac{1}{2}\left[\bar \psi(x) \gamma_\mu \gamma_5 U_\mu(x) \psi(x+a \hat \mu)+
   \bar \psi(x+a \hat \mu) \gamma_\mu \gamma_5 U_\mu^\dagger(x) \psi(x) \right]\,,
   \nonumber
   \\
  P&=&\bar \psi(x) \gamma_5  \psi(x)  \,, \quad   a O_C= \bar \psi(x) \gamma_5 C_{xx} \psi(x)\,,
\\
a \, O_W&=& 8 P - \frac{1}{2}\sum_\mu\left[\bar \psi(x) \gamma_5  U_\mu(x) \psi(x+a\hat \mu)+
   \bar \psi(x+a\hat \mu) \gamma_5  U_\mu^\dagger(x) \psi(x) + ( x \rightarrow  x-a\hat\mu)\right] \,.
   \nonumber
\end{eqnarray}
A possible gauge field smearing has to match that in the fermion matrix of the used action: for SLiNC fermions 
that means stout-smeared $U$s in $\partial \cdot A$ and $O_W$, unsmeared $U$s in $O_C$.

The first term on the r.h.s. in~(\ref{eq:latWI}) has the form of the usual axial Ward identity, the next term is a contact term needed 
(even in the continuum) to make the WI hold offshell.
To give an interpretation of the remaining pieces, consider first the free case (tree level). The clover term vanishes, 
whereas the Wilson term
gives a delta function in position space to cancel lattice artefacts in the propagators~\cite{Capitani:2000xi}.
In the interacting case, $O_C$ and $O_W$ can mix with lower-dimensional operators to make sure that the divergence of 
$A^{\rm split}$ vanishes at $\kappa_c$ instead of $\kappa=1/8$.

In the forward case $G_{jxi}[\partial \cdot A]=0$, so
we are left with an identity linking the propagators to the three-point functions of
$P$, $O_W$, $O_C$. This
could be used to find $\kappa_c$ measuring the three-point functions at different $\kappa$ on-shell.

\section{Perturbative checks and renormalisation}
\vspace{-2mm}

For the check of the axial WI in one-loop lattice perturbation theory (LPT) including partly $O(a^2)$ lattice corrections
in momentum space with initial (final) quark momentum $p_1$($p_2$) [$\Lambda_O=\Lambda_O(p_2,p_1)$]
\begin{eqnarray}
  \gamma_5 S^{-1}(p_1)+S^{-1} (p_2)\gamma_5= - \Lambda_{\partial \cdot A} + 2 m_B\Lambda_P+ 2 \Lambda_C+\Lambda_W
  \label{eq:pertWI}
\end{eqnarray}
we have used the Wilson gauge action and clover fermions with stout smeared $U$ in $\partial \cdot A$ and $O_W$ and 
unsmeared $U$ in $O_C$, the $\Lambda$s denote amputated Green functions.
In perturbation theory
\begin{eqnarray}
   am_B= am - \frac{C_F g^2}{16 \pi^2} \Sigma_0 \,, \quad C_F=\frac{4}{3} \,, 
\end{eqnarray}
where $\Sigma_0$ is the one-loop $1/a$ contribution to the quark self energy.

The needed Feynman rules have been derived, in the forward case ($p_2=p_1$) one-loop calculations have been performed
for arbitrary quark masses including $O(a^2)$ effects. Our results have been checked against results 
from the Cyprus group~\cite{Constantinou:2009tr} where available. However, our technique to include $O(a^2)$ differs and avoids Monte Carlo 
estimates of integrals. In the nonforward case so far a mass expansion to $O(am)$ has been considered.

\begin{figure}[!htb]
  \begin{center}
     \includegraphics[scale=0.6,clip=true]{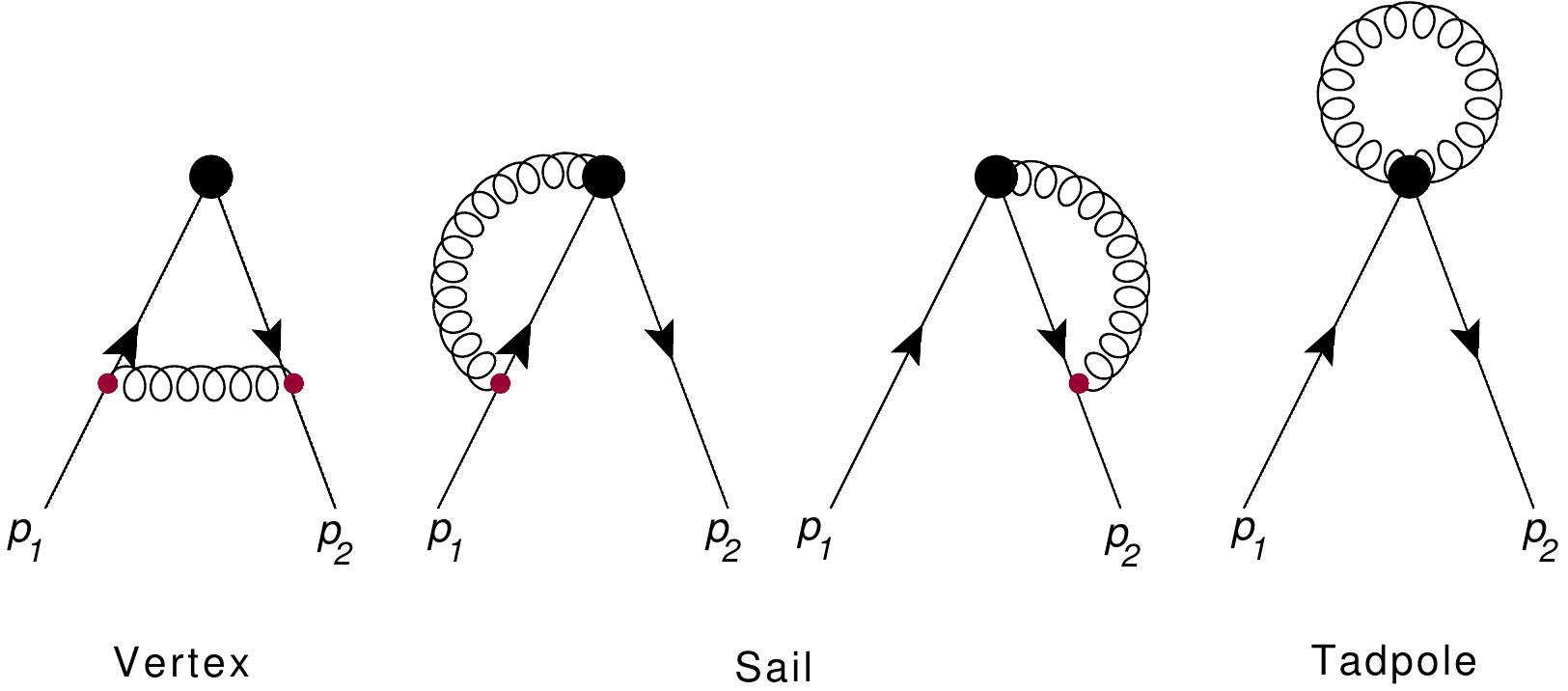}
  \end{center}
  \vspace{-3mm}
  \caption{Typical diagrams.}
  \label{fig:diag}
\end{figure}

Typical diagrams are shown in Fig.~\ref{fig:diag}.
Vertex, sail and tadpole diagrams have to be taken into account for the operators $A^{\rm split}$, 
$\partial \cdot A$ and $O_W$, 
to $P$ only the vertex diagram  and to $O_C$ the sail diagrams contribute.
The results are functions of $p_1, p_2, m,\alpha, \csw, \omega$, 
where $\alpha$ is the gauge parameter ($\alpha=1$ Feynman gauge), $\csw$
the strength of the clover term and $\omega$ the link smearing parameter (for further notations see~\cite{Horsley:2008ap}).

We verified the known loop contribution to $\kappa_c$ from the mass independent contributions of~(\ref{eq:pertWI}) 
both in the forward and nonforward case.

The renormalisation factors (in $\MS$) are defined as usual from the quark propagator and the amputated Green's functions
\begin{eqnarray}
  S^\MS(p,m^\MS) =  { { Z_2^{-1}}}\,  S^{\rm lat}(p,{ {Z_m}} m)\,,
  \quad
  \Lambda_O^\MS (p_2,p_1) =  {Z_O} \, \Lambda_O^{\rm lat}(p_2,p_1) \,.
\end{eqnarray}
$Z_2$, $Z_m$, $Z_P$ and $Z_{A^{\rm loc}}$ agree with the known results~\cite{Constantinou:2009tr}.
New for the used action is the renormalisation factor for the point-split axial vector current
(in $\MS$)
\begin{eqnarray}
  && { { Z_{A^{\rm split}} =Z_{\partial \cdot A}=
   1+ \displaystyle{\frac{g^2 C_F}{16 \pi^2}}
  \Big[
    -8.66279 + 0.9116267 \csw +}}
  \\
  && \hspace{+2mm} { {2.00070 \csw^2  + 85.9927 \omega
   - 4.77316 \csw \omega - 282.301 \omega^2 \Big]}}
  \,.
  \nonumber
\end{eqnarray}
In the continuum from the axial WI one finds $Z_m Z_P = Z_A$.
As expected, for the naive lattice axial WI that equality is not fulfilled (both for the local $A^{\rm loc}$ and the point-split $A^{\rm split}$), the difference
$Z_m Z_P - Z_A$ contains $O(g^2)$ contributions.

Studying the improved identity (\ref{eq:pertWI}) in the forward case, the l.h.s. containing the propagators
is multiplicatively renormalised by $Z_2$
\begin{eqnarray}
  (\gamma_5 S^{-1}+S^{-1}\gamma_5)^\MS = Z_2  (\gamma_5 S^{-1}+S^{-1}\gamma_5)^{\rm lat} \,.
\end{eqnarray}
We checked explicitly in one-loop LPT that
the combined operator on the r.h.s. 
\begin{eqnarray}
2 m_B \bar P \equiv 2 m_B P + 2 O_C + O_W
\end{eqnarray}
is renormalisation invariant
\begin{eqnarray}
  (2 m \Lambda_P)^{\MS} = Z_2 ( 2 m_B  \Lambda_{\bar P}) ^{\rm lat} \,.
\end{eqnarray} 
The local operator $2 m P$ would need a finite renormalisation $Z_m Z_P$ whereas the 
operator with $m_B$ and "irrelevant" 
terms needs neither additive nor multiplicative renormalisation.

In the nonforward case the axial WI  contains the operator $\partial \cdot A$ with the point-split axial vector.
The result of that one-loop study is
\begin{eqnarray}
  \left( -\partial \cdot A + 2 m P\right)^\MS=
  \left( -\partial \cdot A + 2 m_B \bar P\right)^{\rm lat} 
\end{eqnarray}
which shows that the constructed axial WI is renormalisation invariant.

We identify the form of the renormalisation mixing matrix as
\begin{eqnarray}
  \left(\begin{array}{c} -\partial \cdot A \\ 
                           2 m  P \end{array}  \right)^\MS =
  { \left(\begin{array}{cc} Z_{A^{\rm split}} & 0 \\
                          1-Z_{A^{\rm split}} & 1 \end{array} \right) }
  \left(\begin{array}{c} - \partial \cdot A\\
                           2 m_B \bar P  \end{array}  \right)^{\rm latt}
                            \,.
\end{eqnarray}
So the proposed lattice axial Ward identity passed all tests in one-loop LPT.

\section{Nonperturbative check and a first application}
\vspace{-2mm}

We have also started a nonperturbative check of the proposed axial WI.
As action we use the $n_f=2+1$ Stout Link Nonperturbative Clover or SLiNC action~\cite{Cundy:2009yy}.
For the present study we used as parameters $32^3\times 64$, $\csw=2.65$, $\omega=0.1$, 
$\beta=5.5$ [$a=0.074(2)$ fm] and  analysed a subset of available configurations
starting from the flavor symmetric point along a line of constant single quark mass
corresponding to pion masses 465, 360 and 310 MeV [$(\kappa_l, \kappa_s)=(0.120900,0.120900)$, $(0.121040,0.120620)$ 
and $(0.121095,0.120512)$].
We have measured two-point correlation functions $C_{\pi O}(t)$ as function of a ``time'' $t$,
where as source we used a smeared pion at rest and as sink at $t$ we used as $O$ either a smeared pion or
the operators appearing in the axial WI. Since sink and source are time separated we do not 
have contributions from ``contact'' terms.

\begin{figure}
  \begin{center}
  \begin{tabular}{cc}
     \includegraphics[scale=0.48,clip=true]{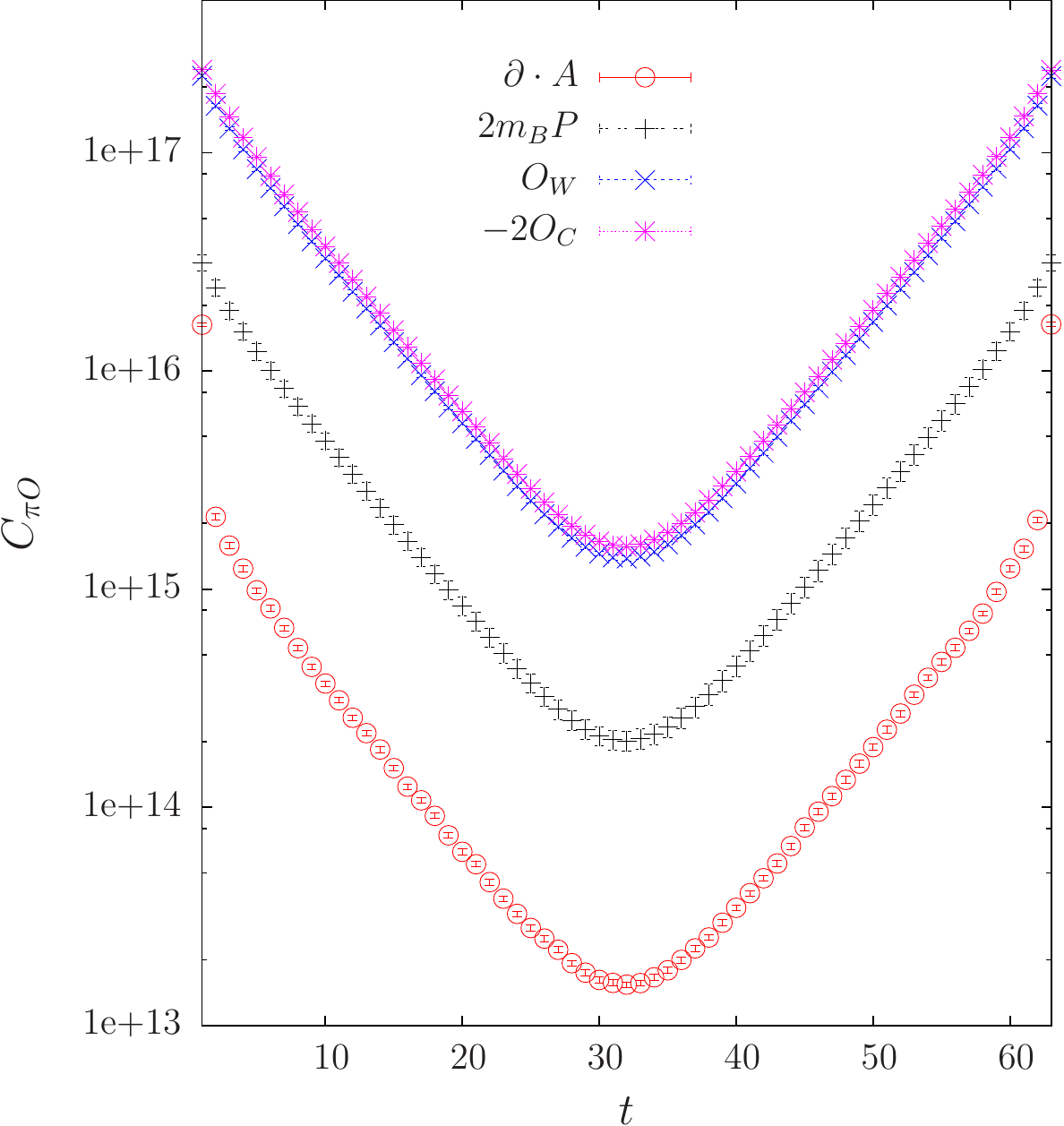}
     &
     \includegraphics[scale=0.48,clip=true]{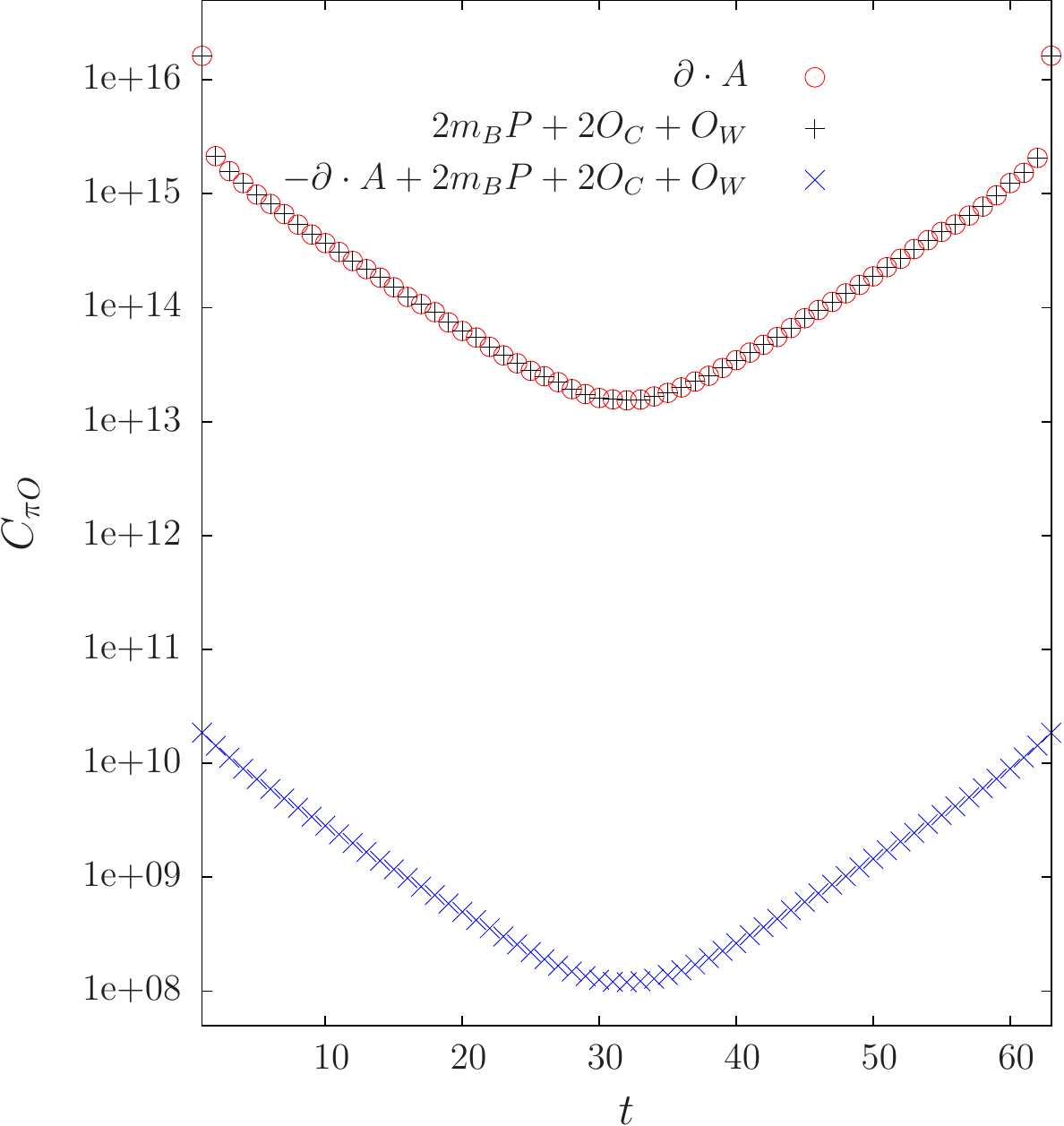}
  \end{tabular}
  \end{center}
  \vspace{-3mm}
  \caption{Correlation functions: (l) individual, (r) combinations at the flavor symmetric point as function of the ``time'' separation $t$.}
  \label{fig:corr}
\end{figure}

Individual correlation functions and several combinations of them are shown in Fig.~\ref{fig:corr} 
showing nicely the cancellation.

\begin{figure}
  \begin{center}
  \begin{tabular}{cc}
     \includegraphics[scale=0.48,clip=true]{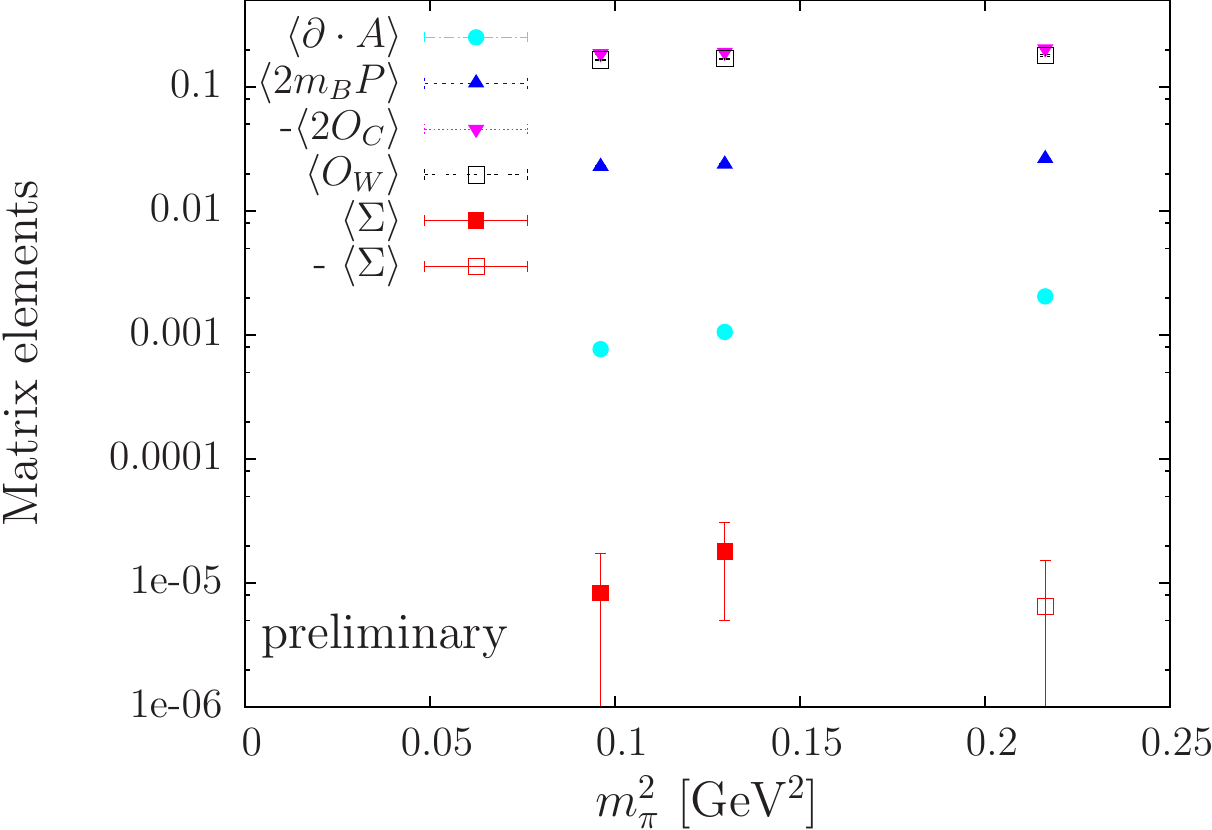} &
     \includegraphics[scale=0.48,clip=true]{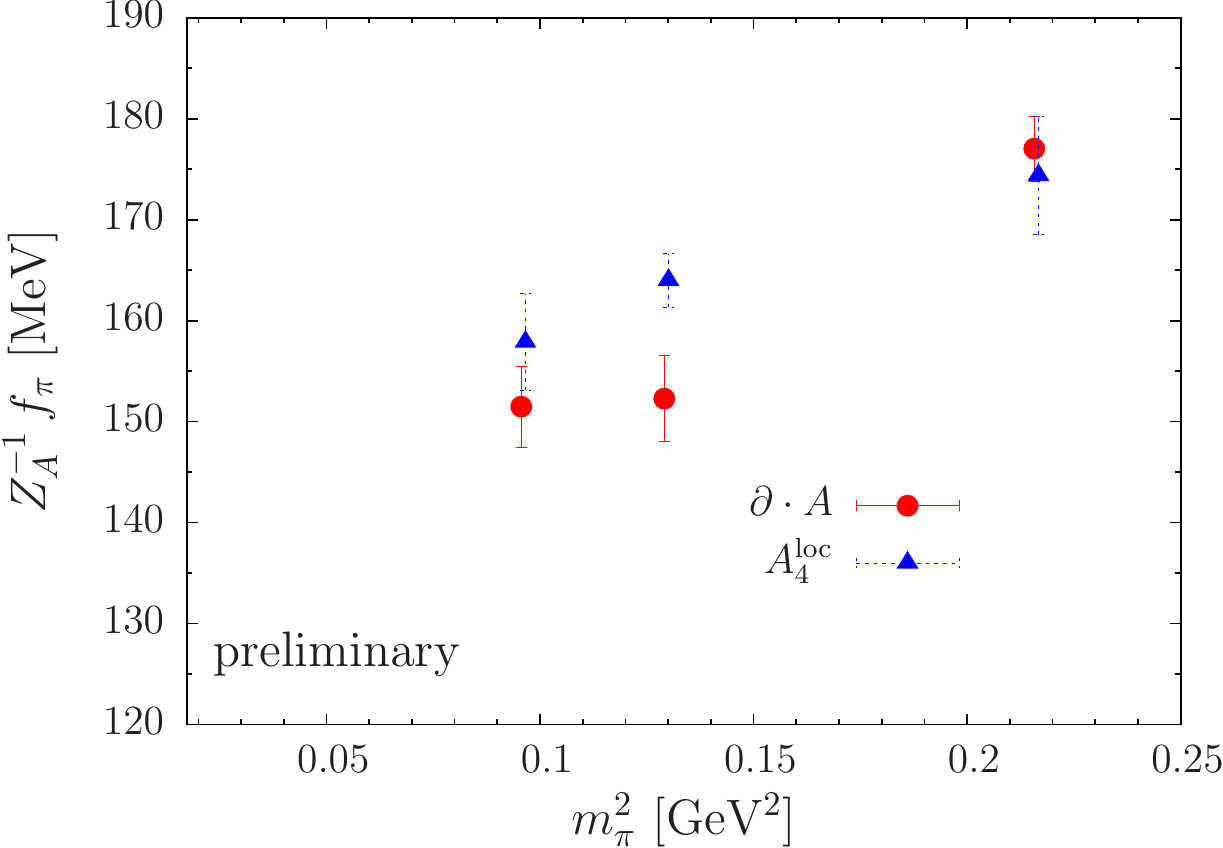}
   \end{tabular}
   \end{center}
  \vspace{-3mm}
    \caption[.]{(l) The individual amplitudes and the check of the WI in (\ref{eq:amp}) for different $m_\pi$. 
              (r) Unrenormalised pion decay constant as function of $m_\pi^2$ using the divergence 
                  of the point-split axial vector current and the local axial vector current.}
   \label{fig:3}           
\end{figure}

The correlators are fitted separately to get the different amplitudes, and then
the amplitudes are combined at bootstrap level to obtain bootstrap errors.
In the left of Fig.~\ref{fig:3}
we present individual averaged amplitudes and
\begin{eqnarray}
  \langle \Sigma\rangle=\langle -\partial \cdot A + 2 m_B P + 2 O_C+ O_W \rangle
  \label{eq:amp}
\end{eqnarray}
as measure how accurate the lattice axial WI is fulfilled (using only nonzero distances in the fit range)
at the different pion masses.
We conclude that the WI passes the nonperturbative check with amazing precision.

The amplitude $\langle \partial \cdot A\rangle$ with the point-split axial vector current is related to the pion decay constant
$f_\pi$.
The unrenormalised $f_\pi$ is obtained using the standard technique of ratios of two-point functions to remove the smeared
pseudoscalar source operator.
In the right of Fig.~\ref{fig:3} we present the unrenormalised $f_\pi$ in MeV derived from $\partial \cdot A$ 
as sink operator with the point-split axial vector current
for a subset of available configurations at the three pion masses. The results are compared to a measurement 
using the local vector current $A_4^{\rm loc}$ as sink with higher statistics.
The results roughly coincide with a tendency of a smaller $f_\pi$ from the point-split current at smaller pion masses.
A final comparison will been performed using the same number of available configurations.

Determining the renormalised pion decay constant from the local axial vector current a renormalisation factor 
$Z_{A^{\rm loc}} =0.873$ has to be used~\cite{Constantinou:2014fka}.
There is the hope that the corresponding nonperturbative Z factor for the point-split axial vector current 
is close to one what is true in one-loop LPT using the Wilson gauge action and clover fermions with stout smeared links.

\section{Short summary and outlook}
\vspace{-2mm}

We have proposed to use a point-split or one-link axial vector present in a lattice axial
Ward identity for clover fermions (\ref{eq:latWI})
instead of the more standard local axial vector violating the naive axial WI with only local operators.
The hope is to improve the accuracy  in the measurements of some hadronic properties using lattice methods.

We have checked in one-loop LPT that this WI is exact in one-loop LPT [partly up to $O(a^2)$] for those fermions.
The renormalisation factor for the point-split axial vector current has been determined for Wilson gauge action
and clover fermions. We have discussed how the renormalisation invariance of that WI is realised.

A first nonperturbative check has been presented using correlators of a pion at rest with the operators
appearing in the WI. For our used lattice parameters at different pion masses the axial WI is fulfilled 
to a very good accuracy.

As a first application we started to determine the (unrenormalised) pion decay constant using the point-split axial vector current.

In the future we plan to check this axial WI at still lighter pion masses and other $\beta$'s
and to measure the renormalised pion decay constant from the divergence of the point-split axial vector.
For that is is necessary to find $Z_{A^{\rm split}}$ nonperturbatively.

Further applications might be alternative $\kappa_c$ determinations and the study of problems with nonzero momentum transfer
(formfactors).

\section*{Acknowledgements}
\vspace{-2mm}
The numerical configuration generation (using the BQCD lattice QCD program~\cite{Nakamura:2010qh}) and data
analysis (using the Chroma software library~\cite{Edwards:2004sx}) was carried out 
on the IBM BlueGene/Q using DIRAC 2 resources (EPCC, Edinburgh, UK),
the BlueGene/P and Q at NIC (J\"ulich, Germany) and the SGI ICE 8200 and Cray XC30 at HLRN 
(Berlin-Hannover, Germany).
HP was supported by DFG Grant No. SCHI 422/10-1. PELR was supported in
part by the STFC under contract ST/G00062X/1 and JMZ was supported by
the Australian Research Council Grant No. FT100100005 and DP140103067.
We thank all funding agencies.

\end{document}